\documentclass{elsart}
\usepackage{epsfig}

\begin{document}

\begin{frontmatter}
\title{Progress of the Volume FEL (VFEL) experiments in millimeter range \thanksref{beltex}}
\footnote[1]{This work is carried out with financial support of private joint-stock company 
BelTechExport, Belarus}

\author{V.G.Baryshevsky}
\author{K.G.Batrakov,}
\author{A.A.Gurinovich,}
\author{I.I.Ilienko,}
\author{A.S.Lobko,}
\author{P.V.Molchanov,}
\author{V.I.Moroz,}
\author{P.F.Sofronov,}
\author{V.I.Stolyarsky.}
\address{Research Institute of Nuclear Problems, 
Belarussian State University,   \\   
11 Bobruyskaya Str., Minsk 220050, Belarus}

\begin{abstract}
Use of non one-dimensional distributed feedback 
in VFEL
gives possibility of frequency tuning in wide range. 
In present 
work dependence of 
lasing process on the angle between resonant diffraction grating grooves and direction 
of electron beam velocity 
is discussed.
\end{abstract}

\end{frontmatter}

\section{Introduction}
New advances in different areas require the
development of tunable, wide-band, high-power sources of coherent
electromagnetic radiation in GigaHertz, TeraHertz and
higher frequency ranges. 
Conventional electron vacuum devices have restricted possibility of
frequency tuning (usually it does not exceed
5-10\%) for the certain carrier frequency at certain e-beam 
energy. Volume free
electron laser (VFEL)\cite{LANL98,LANL2001} was proposed as a new 
type of free electron laser.
Frequency tuning, possibility of use of wide electron beams (several e-beams) 
and reduction of threshold current density 
necessary for start of generation, provided by VFEL, 
make it a basis for 
development of more compact, high-power and tunable radiation sources then 
conventional electron vacuum devices
could let.

First lasing of VFEL
was reported at FEL 2001 \cite{LANL2001}.

\section{Volume FEL distinctive features}
Benefits given by VFEL:

1. Volume FEL provides frequency tuning by rotation of diffraction grating;
\begin{figure}[h]
\epsfxsize = 16 cm 
\centerline{\epsfbox{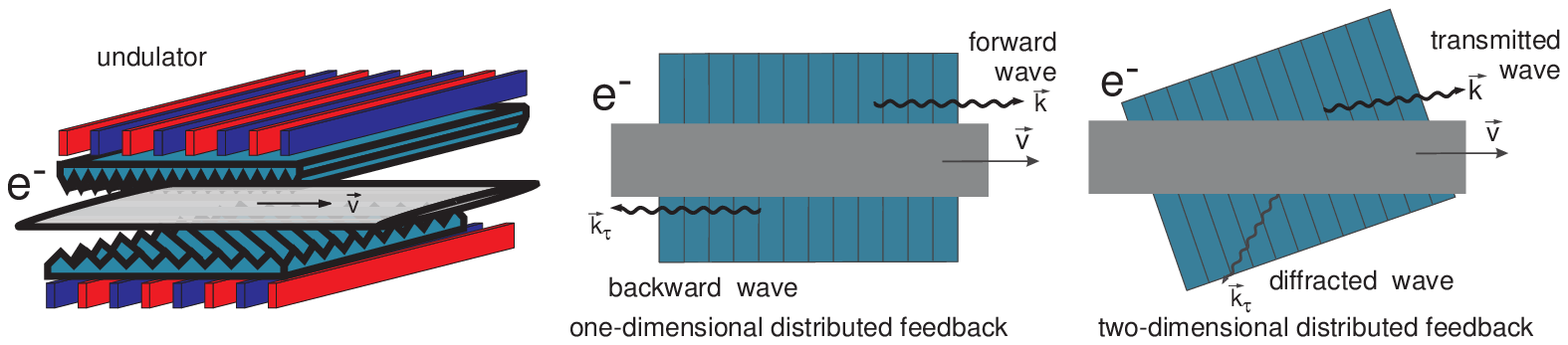}}
\caption{}
\label{rotation}
\end{figure}

2. Use of multi-wave diffraction reduces generation threshold and 
the size of generation zone. 
Starting current j depends on interaction length L as \cite{bar1}: 
$j_{start}\sim 1/\{(kL)^{3}(k\chi _{\tau }L)^{2s}\}$,
$s$ is the number of surplus waves appearing due to diffraction 
(for example, in case of two-wave Bragg diffraction $s=1$, 
for three-wave diffraction $s=2$ and so on).
\begin{figure}[h]
\epsfxsize = 15 cm 
\centerline{\epsfbox{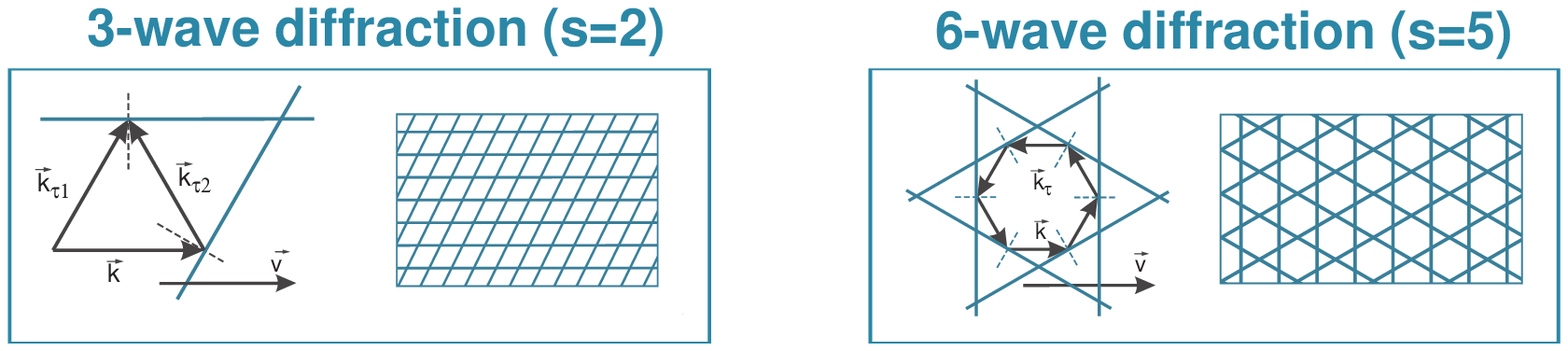}}
\caption{}
\label{multiwave}
\end{figure}  

3. Wide electron beams and diffraction gratings of 
large volumes can be used in VFEL. 
Two or three-dimensional diffraction gratings allow to distribute 
interaction over large 
volume and to overcome power restrictions in resonator.
Volume distributed feedback provides mode discrimination in VFEL resonator.
\begin{figure}[h]
\epsfysize = 4 cm 
\centerline{\epsfbox{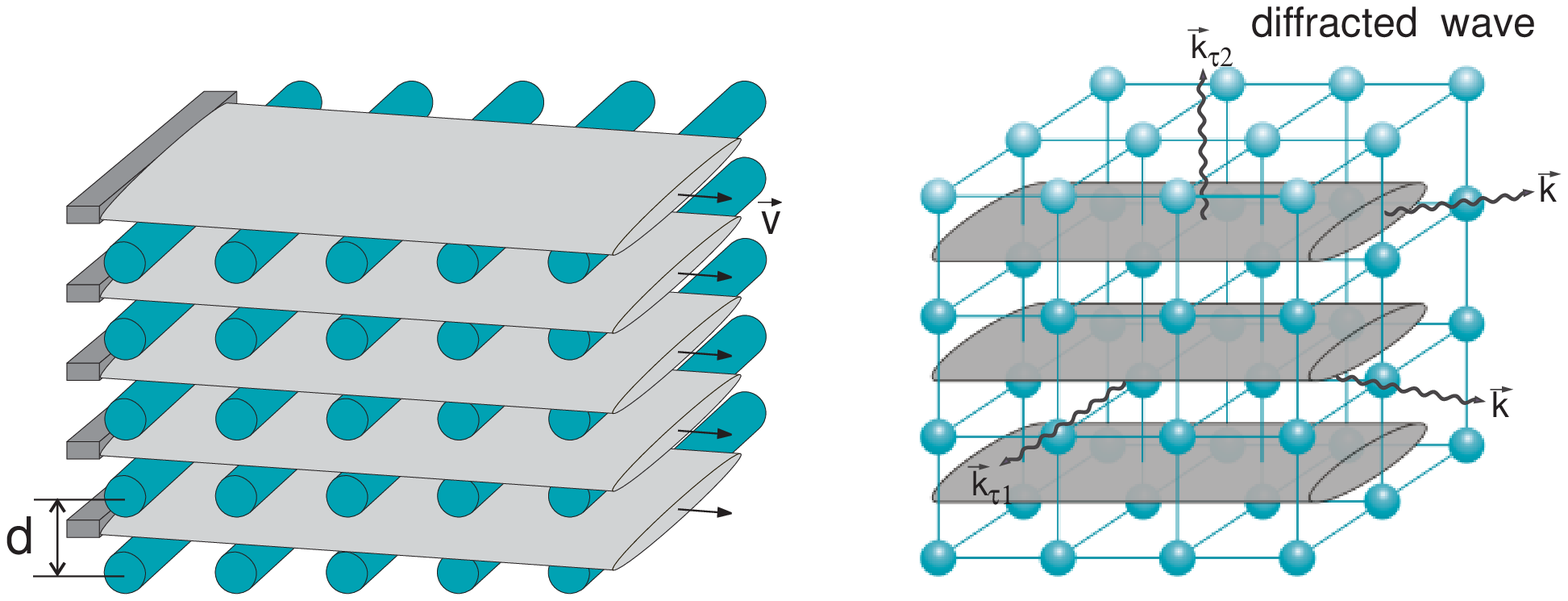}}
\caption{}
\label{volume}
\end{figure}

4. VFEL can simultaneously generate radiation at several frequencies that is extremely 
important, for example, for development of multifrequency radiation source.

\section{Experimental setup}
VFEL operates in wavelength range $\sim$4 mm at electron beam
energy up to 10keV. Such generating device is very
interesting for practical use in submillimeter (TeraHertz)
range, which is actively employing for medical examination. 
VFEL resonator is formed by two parallel diffraction
gratings with different periods and two smooth sidewalls 
(see figure in \cite%
{LANL2001}). The interaction of the exciting diffraction grating
with an electron beam arouses Smith-Purcell radiation.
The resonant diffraction grating provides distributed feedback of generated
radiation with electron beam by Bragg dynamical diffraction. 
Resonant grating can rotate to change orientation
of grating grooves with respect to electron beam velocity that provides
tuning of diffraction conditions. 
\footnote[1]{Undoubtedly, fast frequency modulation of VFEL radiation 
can be done by
rotation of electron beam (for example, by the use of drift in crossed
fields) with respect to diffraction grating grooves. Also frequency tuning
can be provided by azimuth rotation of annular electron beam in resonator
formed by two cylindrical diffraction gratings with helix grooves.
}
\begin{figure}[h]
\epsfxsize = 16 cm 
\centerline{\epsfbox{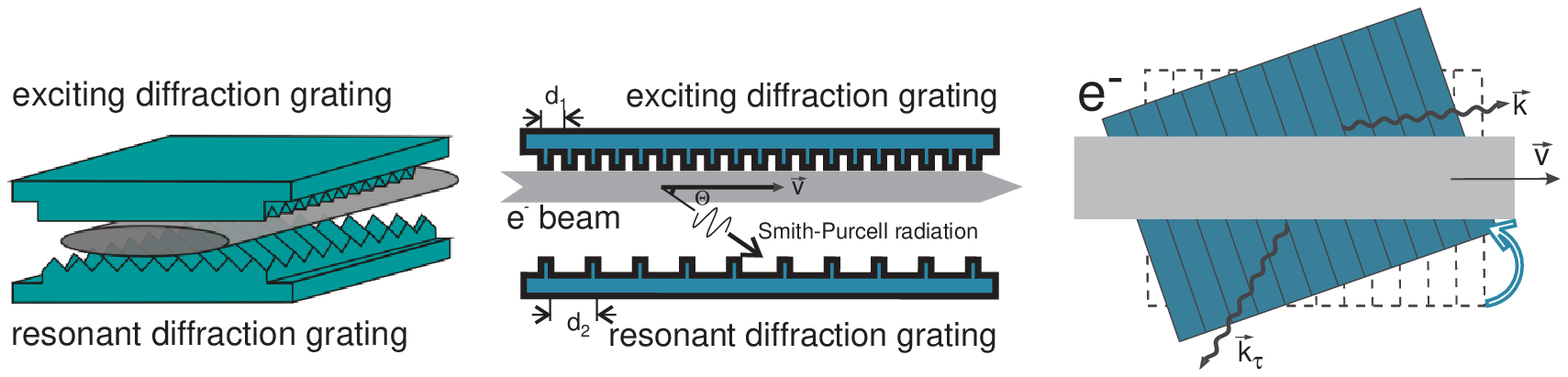}}
\caption{}
\label{resonator}
\end{figure}

The length of resonator is 100 mm,
periods of the exciting diffraction grating and resonant diffraction grating
are 0.67 mm and 3 mm, respectively. The ribbon electron beam with profile 10 mm x 1 mm 
is
emitted by thermal cathode, formed in Pierce gun without beam
compression and guided through resonator along dffraction gratings. Only electrons passing at the distance $h\leq \lambda \beta
\gamma /(4\pi )$ \cite{walsh} near the exciting grating surface interact
with electromagnetic wave effectively (i.e. less than 10\% of total cross
size of electron beam participates in lasing process). To provide the most
effective interaction of electron beam with the exciting grating, it is 
placed in a beam and the value of ``grating current'' (i.e. part
of total electron beam current falling to the grating) is monitored as a
touchstone of interaction efficiency and indicator of beam position with
respect to grating surface. The following parameters can be monitored during
experiment: beam voltage (electron beam energy); total electron beam current
(number of emitted electrons per second); ''grating current''; 
angle between resonant grating grooves and electron beam velocity 
(angle of grating rotation);
distance between gratings (transverse size of resonator); 
microwave power
and frequency. Electrons are emitted in pulsed regime (unipolar pulse with sinusoidal 
shape and pulse duration ~10 ms) in sequence of two or three voltage 
pulses. Voltage can vary from 0 to 10 kV (hence, electron beam energy 
is varied from  0 to 10 keV). Different longitudional modes of 
resonator are excited while electron energy passes through 
the whole variation interval. 
Frequency of data reading is determined by the condition that change of
electron energy between two consecutive measurements of each monitored parameter
should not cause change of generation regime.
Estimated admissible change of
energy is about $\sim$1\% of rating value. Data reading interval is $%
4\cdot 10^{-5}$ s.

A lot of system parameters can be varied during experiment, so complete
investigation of the system requires multiple measurements. 
Experimental study of frequency tuning and change of radiation
intensity at
mechanical rotation of resonant diffraction grating 
(i.e. at variation of the angular orientation of grooves of 
resonant diffraction grating  
with respect to electron beam velocity)
is presented in this paper.

\section{Theoretical background}
General
theoretical statements \cite{PLetters,LANL98} should be reminded 
for better understanding of the below experimental results. 
Considering processes in resonator one should discern the difference
between two cases: (a) presence of resonator sidewalls can be
neglected; (b) presence of sidewalls of resonator can not be neglected.
For unbounded waveguide (case (a)) two geometries of diffraction are distinguished: 
those of Bragg and Laue \cite{cheng}.
Laue geometry of diffraction implies that both incident and diffracted waves
passes to vacuum through the same boundary of resonator, while at
Bragg diffraction incident and diffracted waves passes through the different
boundaries. 
Absolute instability can appear in Bragg geometry and such system
works as generator (for backward Bragg diffraction it converts into
a well-known backward-wave tube).
For Laue case convective instability provides only amplification regime.
\begin{figure}[h]
\epsfysize = 4 cm 
\centerline{\epsfbox{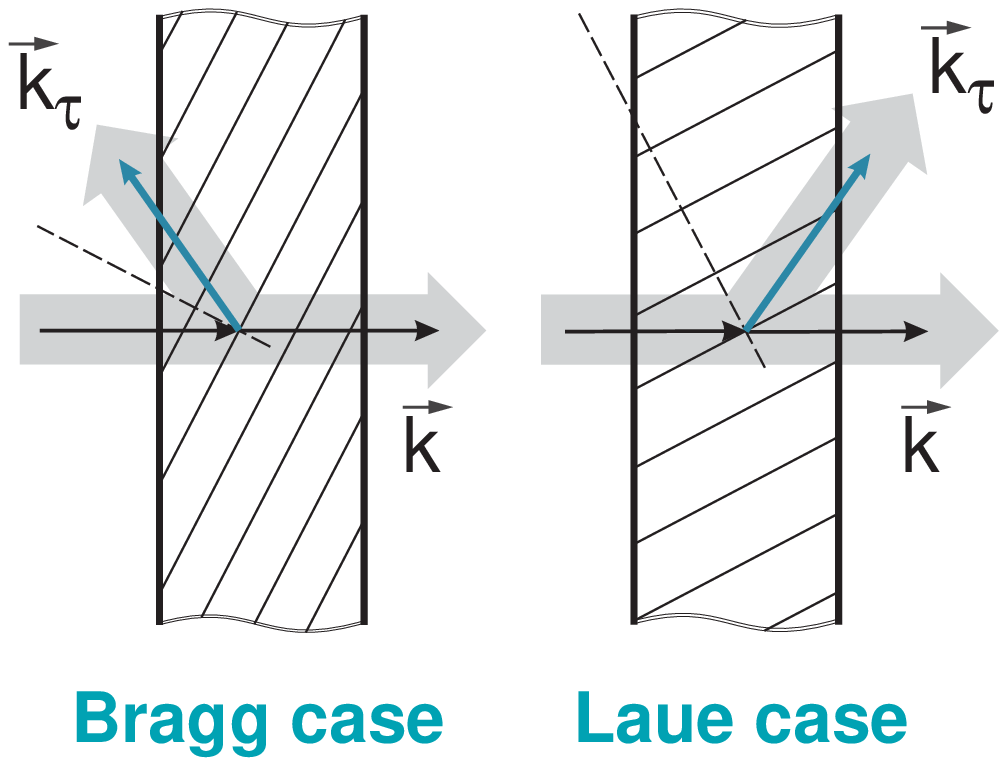}}
\caption{}
\label{bragg}
\end{figure}

Generation process in
resonator 
is described by Maxwell equations, containing 
the space-periodic permittivity of
diffraction grating
$\chi \left( \vec{r},\omega \right) =\sum_{\vec{\tau}\neq 0}\chi
_{\tau }\left( x\right) e^{-i\tau _{y}y}e^{-i\tau _{z}z}$, where 
$\vec{\tau}$ is the reciprocal lattice vector of the diffraction
grating \cite{LANL98}.

If $\chi _{\tau }$ in the above equations is equal to zero, they describe
smooth waveguide, which eigenfunctions $\left| \vec{Y}\left(
x,y\right) \right\rangle$ are well known.
If presence of waveguide sidewalls can not be neglected,
($ \left| \vec{Y}%
_{nm}\left( x,y\right) \right\rangle \sim\sin \frac{\pi n}{a}%
x\,\,\sin \frac{\pi m}{b}y$, where
$n$, $m$ are the integer numbers, $a$ and $b$ are the 
transversal dimensions of resonator).
\begin{figure}[h]
\epsfxsize = 5 cm 
\centerline{\epsfbox{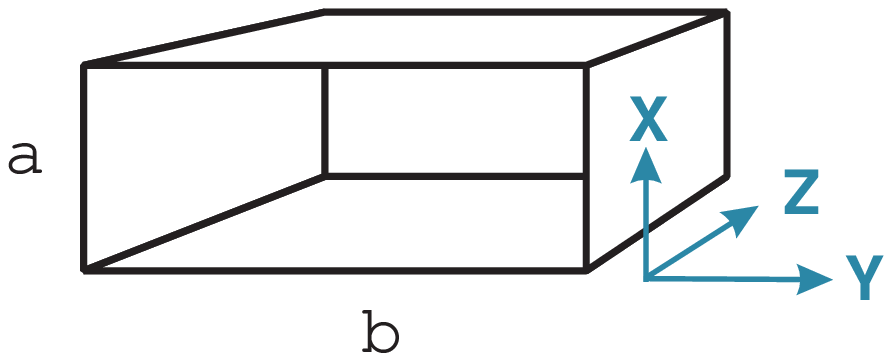}}
\caption{}
\label{waveguide}
\end{figure}

Expanding Maxwell equations over eigenfunctions of smooth waveguide 
$\left| \vec{Y}_{nm}\left( x,y\right) \right\rangle$ one obtains equations
equations describing
generation in continuous effective grating with periodic permittivity
$\chi _{eff}(z)$ \cite{LANL98}.

In single-mode approximation the above problem comes to that of
generation (amplification) in one-dimensional periodic medium with effective
permittivity 
\begin{equation}
\chi _{eff}\left( z\right) =\chi ^{nn(mm)}\left( z\right) \sim \int
dx\,\sin ^{2}\frac{\pi n}{a}x\,\chi _{\tau }\left( x\right) \,\int
dy\,\sin ^{2}\frac{\pi m}{b}y\,e^{-i\tau _{y}y}e^{-i\tau _{z}z}
\label{chi_eff}
\end{equation}
Presence of diffraction grating in waveguide makes nondiagonal
matrix elements $\chi ^{nn^{\prime }(mm^{\prime })}\left( z\right)$, 
describing transition between modes,
different from zero:
\[
\chi ^{nn^{\prime }(mm^{\prime })}\left( z\right) \sim \int
dx\,\sin \frac{\pi n}{a}x\,\chi _{\tau }\left( x\right) \,\sin \frac{\pi
n^{\prime }}{a}x\int dy\,\sin \frac{\pi m}{b}y\,\,e^{-i\tau _{y}y}e^{-i\tau
_{z}z}\sin \frac{\pi m^{\prime }}{b}y.
\]

Presence of
sidewalls imparts new feature to diffraction process in resonator, 
because even in Laue-like case 
(which corresponds 
to Laue geometry of diffraction if neglecting the sidewalls) 
the matrix element 
$\chi _{eff}\left( z\right) $ differs from zero. 
And, as a result, generation regime can be
reached at Laue-like geometry, in contrast with an unbounded waveguide, 
where Laue geometry
provides only amplification regime,
while generation is possible at Bragg diffraction.

\section{Experimental results}
\begin{figure}[h]
\epsfxsize = 16 cm 
\centerline{\epsfbox{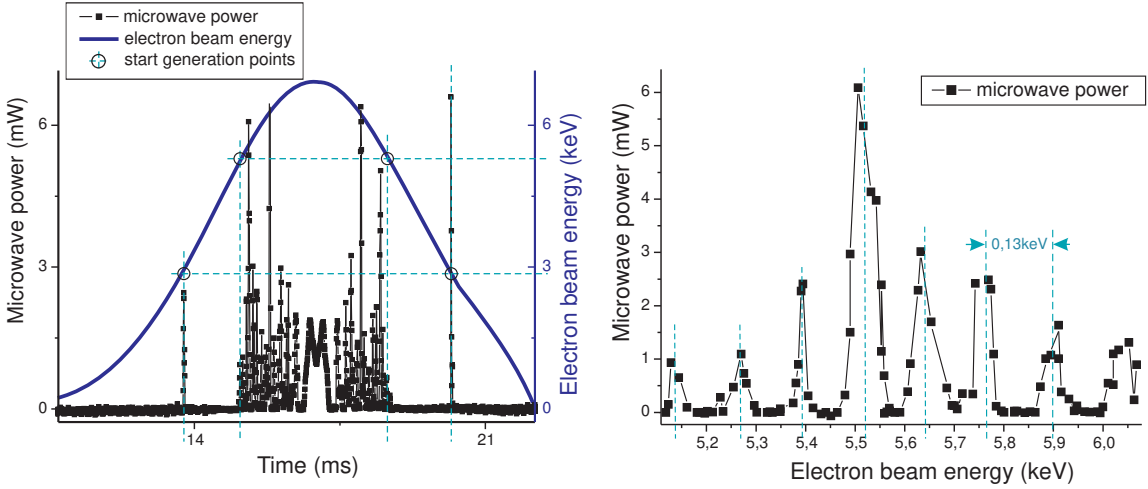}}
\caption{a. General view of microwave signal; b. 
Dependence of microwave power on electron beam energy.}
\label{microwave}
\end{figure}

General view of microwave signal is presented
on Fig.\ref{microwave}a. 
Presence of several microwave peaks is the manifestation of longitudinal cavity modes.
Frequency difference 
$\Delta \nu$ 
associated with the change $\Delta E$ of electron energy $E$  ($%
\nu \sim \sqrt{E(keV)}$) can be estimated as $\Delta
\nu =\nu _{0}\,\frac{\Delta E}{2E_{0}}$.
According to measurements $\nu_0 \approx 54 \,GHz$,
$\Delta E\approx
0,13\,keV$ for two next distinguished peaks (Fig.\ref{microwave}b.), 
hence, frequency difference for them can be estimated as 
$\Delta
\nu \sim ~0,57$ GHz.

\begin{figure}[h]
\epsfysize = 4 cm 
\centerline{\epsfbox{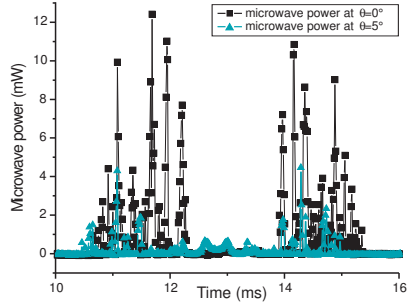}}
\caption{Change of lasing efficiency
with grating rotation, $\theta$ is the angle of grating rotation, 
$\theta=0$ when grating grooves are perpendicular to e-beam velocity.}
\label{voltage}
\end{figure}

Now let us consider behavior of lasing efficiency
with grating rotation. Rotation of diffraction grating changes 
components of
reciprocal lattice vector. From (\ref{chi_eff}) it is easy to 
see that $\chi
_{eff}\left( z\right) $ decreases with $\tau _{y}$ growth.
Change of $\chi _{eff}\left( z\right) $ yields change of generation
conditions and, in particular, generation threshold. Reduction of $\chi
_{eff}\left( z\right) $ results in decrease of generation efficiency 
(Fig.\ref{voltage}). 
But,
interaction efficiency can be increased if recollect that at
certain value of current density the intensity is proportional to 
the term $%
e^{-\frac{4\pi }{\lambda \beta \gamma }h}$\cite{walsh} where $h$ 
is the
distance from electron to the grating, $\gamma $ is the Lorentz-factor, 
$\lambda $ is the radiation wavelength, $\beta =v/c=\sqrt{2E(keV)/511}$. 
As a result, energy of electron beam at certain current 
density should be
increased (to increase factor $e^{-\frac{4\pi }{\lambda \beta \gamma }h}$)
to overcome decrease of $\chi _{eff}$.

\begin{figure}[h]
\epsfysize = 4 cm 
\centerline{\epsfbox{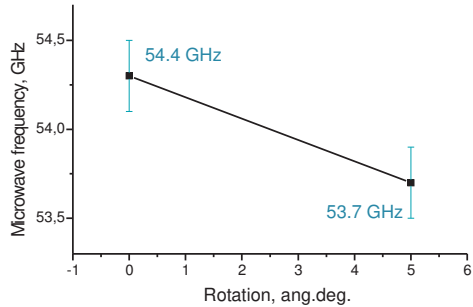}}
\caption{Measured frequencies at rotation of resonant diffraction grating. }
\label{frequency}
\end{figure}

Non-one-dimensional feedback, being used in VFEL, 
provides tuning of radiation frequency
by diffraction grating rotation. 
Measured frequencies at rotation of resonant diffraction grating 
for microwave peak, corresponding electron beam energy 2,9 keV are 
shown in Fig.\ref{frequency}. Measured frequency change well accords
theoretical predictions \cite{PLetters,LANL98}

\section{Conclusion}
VFEL generator can be considered as a new type of tunable backward wave tube with
variable period (due to gratings rotation) and VFEL amplifier is a new type of
tunable travelling wave tube.

Frequency tuning, possibility of use of wide electron beams (several e-beams) 
and reduction of threshold current density 
necessary for start of generation, provided by VFEL, 
make it a basis for 
development of more compact, high-power and tunable radiation sources then 
conventional electron vacuum devices
could let.

New VFELs with electron beam energy up to 30keV and 500keV 
have been developing 
for next set of experiments.


\end{document}